\documentclass{article}

\usepackage[utf8]{inputenc}
\usepackage{amsthm,amsmath,amssymb}
\usepackage{graphicx}
\usepackage{authblk,setspace}
\usepackage[shortlabels]{enumitem}
\usepackage[a4paper, total={6in, 8in}]{geometry}
\usepackage{lscape}
\usepackage{booktabs}
\usepackage{moreverb,url,setspace}
\usepackage{moreverb}
\usepackage[numbers]{natbib}
\usepackage{subfigure}
\usepackage{subcaption}

\date{}

\title{Evaluating the impact of outcome delay on the efficiency of sample size re-estimation}

\author[1]{Aritra Mukherjee}
\author[2]{Michael J. Grayling}
\author[1]{James M. S. Wason}

\affil[1]{Population Health Sciences Institute, Newcastle University}
\affil[2]{Janssen R\&D}

\begin{document}
\maketitle

\abstract{

Sample size re-estimation (SSR) can be a powerful tool to ensure that a clinical trial meets its pre-specified power requirements when uncertainty regarding a design parameter exists at the planning stage. 
However, long-term primary endpoints can be harmful to the efficiency of this trial design. 
If recruitment is continued while treatment outcomes are awaited, long delay can potentially lead to a large number of `pipeline' participants being recruited in the trial that do not contribute to the interim analysis.
This may lead to a larger number of recruited participants than are actually deemed required, resulting in an over-powered trial with high cost.
This paper studies the exact impact of such outcome delay on the efficiency of `internal pilot' type SSR designs.
The distribution of the final sample size post SSR is obtained under various delay lengths for both continuous and binary outcome data; how delay impacts the precision of the final sample size estimate is then discussed.
Precisely, the impact of delay on this precision is assessed through RMSE, as well as two more novel metrics, termed the \textit{delay impact} and \textit{cost}.
The results indicate that with increase in delay length, the \textit{delay impact} increases, inflating average sample size and power.
However, the severity of the effect of delayed outcomes depends highly on the exact trial setting.
Trials where the re-estimated sample size is smaller than originally planned suffer the most from delayed outcomes, often leading to an over-powered trial.
However, the impact of delay is substantially less if the original planned sample size remains smaller than the re-estimated sample size.\\


}

\section{Introduction}

Sample size estimation is an integral part of every clinical trial.
This is because it is important to be able to detect a pre-specified treatment effect with the correct power, such that efficacious drugs have a high chance of being identified without using unnecessary resources.
The estimation of sample size requires estimates of nuisance parameter(s) along with the treatment effect.
These nuisance parameters could reflect, e.g., the outcome variance, and intra-class correlation coefficient for clustered settings, or the population event rate, depending on the type of data and the study design. 
Unfortunately, in practice, there is often little information available on these nuisance parameters at the trial planning stage.
Similarly, a particular treatment effect may be assumed in a sample size calculation that poorly reflects the true population treatment effect; this is problematic as the trial will be incorrectly powered.
In such scenarios, a sample size re-estimation (SSR) design may be useful \cite{Chang2014}. 

A SSR design allows adjustment of the sample size of the trial based on accrued participant data on nuisance parameter(s) and/or the treatment effect, in order to achieve a pre-specified power level.
There are several available approaches \cite{Friede2004,Friede2006,Friede2013,Kieser2003,Gao2008,Proschan2009,Shih2016,Wang2021} to SSR, often sub-classified as to whether they are blinded or unblinded. 
As the name suggests, blinded SSR preserves the blinding of participant allocations to the treatment arms.
Whereas, in unblinded SSR, the treatment allocation is revealed at the interim: often this will be because an estimate of the treatment effect is desired for use in a conditional power calculation \cite{Kunzmann2022,Gang2002,Jennison2015,Edwards2020}.
Where possible, blinded SSR is preferred over unblinded SSR, for better preservation of the integrity of the trial.

While the literature suggests many different approaches for SSR, a common assumption across these articles is that the treatment outcomes are immediately available. 
This might poorly reflect many trials in practice.
In the presence of such outcome delay, the trial might stop recruitment while awaiting treatment outcomes during the interim analysis, prolonging the trial duration. 
Or, the trial might continue to recruit participants during the interim analysis leading to a number of \textit{`pipeline'} participants being recruited in the trial for whom we do not have the treatment outcome available.
Either of the above scenarios is not desirable for an efficient trial \cite{wason2019,MukherjeeGrayling2022}. 
While there exists methods for SSR which include partial information from participants\cite{Wust2003,Wust2005}, these are rarely taken up in practice.
Earlier studies on Simon's two-stage design \cite{Mukherjee2022}, two-arm group-sequential designs \cite{Mukherjee2025GSD} and multi-arm multi-stage designs \cite{Mukherjee2025MAMS} showed that a delayed outcome heavily impacts the efficiency of these trial designs. 
While Wason et al. considered SSR less susceptible to outcome delay than other adaptive designs, it was noted further research was required \cite{wason2019}.
Indeed, the influence of such pipeline participants has been significant on the expected sample sizes of multi-stage trials with futility/efficacy interim assessments, but there has been no study that summarizes how a delayed outcome impacts other kinds of adaptation.
Therefore, in this paper, we aim to analyse the impact of long-term primary outcomes on the efficiency of SSR designs. 
We focus on the often used `internal pilot' type of SSR design, that does not re-estimate the treatment effect assumed in the sample size calculation and thus can typically be performed in a blinded manner.
In particular, the distribution of the re-estimated sample size is used to assess the impact of outcome delay through a number of summary metrics.

\section{Motivating example}

As an example, consider the phase III randomized placebo-controlled trial (NCT02836496) that assessed the efficacy of mepolizumab for hypereosinophilic syndrome \cite{Roufosse2020}.
In this trial, the primary outcome was the proportion of participants who experienced a hyper eosinophilic syndrome flare during a 32 week study period.
Participants were recruited from March 7th 2017 until October 18th 2018.
Therefore, the total recruitment length was approximately 19 months, with the primary outcome taking 32 weeks to observe following enrollment.
An initial sample size of 80 participants (under 1:1 randomization) was estimated as being required to achieve 90\% power to detect an absolute reduction of 38\% (at a two-sided $\alpha = 0.05$) in the proportion of participants experiencing a flare during the study period.
The initial assumption for the true proportion of participants experiencing a flare on placebo was 60\%. 

Due to a lack of evidence to support this estimate of 60\%, a pre-planned blinded SSR was conducted, with an increase in sample size to be carried out if the blinded overall flare rate was less than 30\%.
The interim analysis was planned after 30 participants were recruited in each arm and the maximum sample size allowed was 120.
Based on the observed data, the re-estimated sample size was set to be 100. 

Per clinicaltrials.gov, the trial recruited a total of 108 participants in 19 months.
Assuming that participants were recruited uniformly over this 19 month period, the average rate of participant recruitment would have been approximately 6 participants/month.
Therefore, if recruitment was not paused during the follow-up period after 60 participants had been recruited, the number of participants that would have been recruited while the primary outcomes were awaited to conduct the interim analysis for 32 weeks would have been approximately 48.
Thus, by the time of the completion of the interim analysis, roughly 108 participants would have been randomised.
However, the re-estimated required sample size was only 100 participants.
Because of the delay in observing the primary outcome, the trial could have recruited more participants than the re-estimation determined were required.
If this time to observing the primary treatment outcome was even larger, the number of extra pipeline participants could have increased further, resulting in a potentially overpowered trial with an increased cost.

\section{Methods}

\subsection{Notation}\label{notation}

We assume an RCT is to be conducted to test the efficacy of an experimental treatment ($E$) against a control ($C$). 
Let $Y_{ij}$ denote the treatment outcome from participant $j = 1, 2, \dots, n_i$ in arm $i = C, E$, and suppose that $Y_{ij} \sim N(\mu_i, \sigma^2)$.
Although we make this distributional restriction, note that the following methods can be readily applied to more general distributions; as an example, in the Supplementary Materials we discuss application to binary outcomes.

We want to test the hypothesis $H_0 : \delta = \mu_E - \mu_C \le 0$ against $H_1 : \delta > 0$, at level $\alpha$ with power $1 - \beta$ when $\delta = \delta_1 > 0$.
As we focus on re-estimation of $\sigma$, we will assume throughout that $\delta = \delta_1$.

We may use an independent two sample $t$-test for the hypothesis test, with test statistic $T$ given as 
$$ T = \frac{\bar{Y}_E - \bar{Y}_C}{s_\text{pooled}\sqrt{\frac{1}{n_E} + \frac{1}{n_C}}}. $$
Here, $s_\text{pooled}^2$ is the pooled sample variance given as
$$ s_\text{pooled}^2 = \frac{(n_E - 1)s_E^2 + (n_C - 1)s_C^2}{n_E + n_C - 2}, $$
where $s_i^2$ denotes the sample variance in treatment arm $i$.

The test statistic $T$ follows a non-central $t$-distribution, with $n_E + n_C - 2$ degrees of freedom and a non-centrality parameter $\nu$, which is a function of $\delta$, $n_E$, $n_C$, and $\sigma$.
The null hypothesis is rejected when $T \ge t_{n_E + n_C - 2}(1 - \alpha)$, the $(1-\alpha)$ quantile of a $t_{n_E + n_C - 2}$ distribution.
For simplicity, we now assume equal sample allocation across the arms, i.e., $n_E = n_C = n/2$.
Then, for the above test, in order to achieve the required power, the following formula is often used based on asymptotic normality to set the required sample size
\begin{equation}\label{eq:ch4:N_single}
n(\sigma) = \frac{4\sigma^2\{\Phi^{-1}(1 - \alpha) + \Phi^{-1}(1 - \beta)\}^2}{\delta_1^2}.
\end{equation}
Here, we list the sample size explicitly as a function of $\sigma$, since we will focus on re-estimating this parameter.
In what follows, we will refer for brevity to the design that happens to know the value of $\sigma$ as the `oracle design', which would use sample size $n_\text{oracle} = n(\sigma)$.

\subsection{Sample size re-estimation}\label{SSR methodology}

At the planning stage of a trial, we typically do not know the value of $\sigma$ and start the trial based on some assumption, say $\sigma = \sigma_{init}$, which we may lack confidence about.
An associated initial estimate of the required sample size is set as $n_{init} = n(\sigma_{init})$.
Then, we assume a SSR design is chosen to estimate $\sigma$ at an interim analysis and adjust the sample size if required to ensure sufficient power for the trial.
Since blinded SSR is often considered to be more preferable \cite{Burton2006}, the study here onwards uses blinded SSR to estimate $\sigma$.
Specifically, the re-estimation is based on a pooled estimate of the sample variance \cite{Kieser2003} given as $s^2_{pooled}$ in the \textit{Notation} section.
Methods have been proposed that reestimate the variance using an EM algorithm assuming the observations from the blinded data comes from a mixture of two normal distributions\cite{GouldShih1992}.
We choose to omit this approach as the estimates thus obtained tend to underestimate the variance substantially \cite{Friede2002,Friede2006}.

With the above, SSR can be looked upon as an estimation problem where we try to estimate $n_\text{oracle}$ as precisely as possible, i.e., the re-estimated sample size is an estimate of $n_\text{oracle}$.

We assume after we observe data from $n_1 < n_{init}$ participants, we estimate the value of $\sigma$, say, $\sigma = \sigma_*$.
Then, SSR is conducted through adjusting the sample size by updating the value of $\sigma$ in equation \ref{eq:ch4:N_single}.
That is, the re-estimated sample size is $n_* = n(\sigma_*)$.
Alternatively, we may write the re-estimated required sample size as $n_* = n_1 + n_{2*}$.

If it takes $m$ units of time to observe the primary outcome data from a participant being randomized, and recruitment is not paused during this delay length at interim analysis, then in the presence of such delay there are two possible cases:
\begin{enumerate}
    \item The delay period $m$ is such that the number of participants recruited during that time, along with the first stage sample size, is smaller than the re-estimated sample size. 
    In this case, delay has no impact on the efficiency of the SSR design, rather, it reduces the time to complete the trial due to continuous recruitment when compared to a trial where we stop recruitment for the interim analysis.
    \item The delay period $m$ is such that the number of participants recruited during the delay period, along with the first stage sample size, is greater than the re-estimated sample size. 
    In contrast to the previous scenario, here the efficiency of the SSR design is impacted; we recruit more participants to the trial than are deemed required. 
\end{enumerate}

Let us denote the number of participants recruited during the $m$ delay period across both arms as $n_\text{delay}$.
We refer to these as `pipeline' participants.
The final total sample size of a SSR design in the presence of delay can then be expressed as
\begin{align}\label{eq:ch4:N*}
    N^* = \begin{cases}
            n_* = n_1 + n_{2*}   &:\ n_{2*} > n_\text{delay}, \\
            n_1 + n_\text{delay} &:\ n_{2*} \le n_\text{delay}.
          \end{cases}
\end{align}

We can estimate the number of recruited participants during the delay period, $n_\text{delay}$, assuming a specified recruitment pattern.
The methods for this are given in the following subsection.

\subsection{Computing the number of pipeline participants}\label{Number of pipelines_SSR}

In this paper, we assume a fixed recruitment pattern that is based on the initial assumption about $\sigma$.
Specifically, we assume that it will take an estimated $t$ units of time to recruit the total $n_{init}$ participants estimated initially in the planning stage, as required based on the assumption $\sigma = \sigma_{init}$.
Further, suppose it takes $t_1$ units of time to recruit the first $n_1$ participants. 
To estimate $n_\text{delay}$, the number of pipeline participants recruited during the $m$ units of time after the $n_1^\text{th}$ participant is recruited, we consider two sub-cases for the recruitment pattern: uniform and linear.
In practice, for a large multi-centre trial, a linearly increasing recruitment pattern is typically observed as new trial sites open.
Once all sites become fully operational, the recruitment rate typically plateaus, with participant arrival following a uniform pattern with a constant recruitment rate.
The two aforementioned sub-cases, of uniform and linear recruitment patterns, serve to represent extremes of this type of recruitment pattern.

\subsubsection{Uniform recruitment}

If participant recruitment follows a uniform pattern during the trial, i.e., we assume a Poisson arrival of participants with parameter $\lambda$, then the best estimate of $\lambda$ is $n_{init}/t$.
Furthermore, $E(n_\text{delay}) = m\lambda$.

\subsubsection{Linear recruitment}

Let us consider an increasing participant recruitment rate such that the recruitment rate across arms is a linear function of time, say $\lambda = \gamma t$, where $\gamma$ is an unknown constant and $t = 1, 2, \dots$.
Then, in $t$ units of time the number of recruitments assuming this trend would be
\[ \gamma(1 + 2 + \cdots + t) = \gamma\frac{t(t + 1)}{2}. \]
As per the assumptions, this value should be equal to the initially planned sample size, $n_{init}$.
Equating this to $n_{init}$ gives an estimate for $\gamma$
\begin{equation} \label{eq:ch4:gamma}
    \gamma = \frac{2n_{init}}{t(t + 1)}.
\end{equation}
Similarly, if we equate the number of recruitments in $t_1$ units of time with $n_1$ participants, we have
$$ \frac{\gamma t_1 (t_1 + 1)}{2} = n_1. $$
Solving this for $t_1$ (restricting to the positive root since time is positive), we get
\begin{equation} \label{eq:ch4:t1}
t_1 = -\frac{1}{2} + \frac{1}{2}\sqrt{1 + \frac{4n_1 t(t + 1)}{n_{init}}}.
\end{equation}
The number of participants recruited after time $t_1$, during the $m$ units of time awaiting the outcome results, is thus
\begin{align*}
    n_\text{delay} &= \gamma[(t_1 + 1) + (t_1 + 2) + \cdots + (t_1 + m)],\\
                   &= \gamma m t_1 + \frac{\gamma m (m + 1)}{2},   
\end{align*}
where values for $\gamma$ and $t_1$ can be acquired from Equations~(\ref{eq:ch4:gamma})-(\ref{eq:ch4:t1}).

Note that as the total time to recruit all $n_{init}$ participants, $t$, has been fixed, this makes $n_\text{delay}$ dependent on the initially planned sample size $n_{init}$ through the recruitment rate assumptions.
In turn, this makes $n_\text{delay}$ dependent on the initial assumptions regarding $\sigma_{init}$, as well as $\delta_0$, $\alpha$, and $\beta$.
Of course, in practice, the (observed) recruitment rate may not be so directly dependent on parameters such as $\sigma_{init}$.
However, it is common practice in trials to choose the number of sites to influence the recruitment rate to limit the planned trial duration to an acceptable length.
Consequently, we believe it is logical to set the recruitment rate for our evaluation in this manner.

Furthermore, we have not set a maximum limit to the sample size in order to not restrict the pipelines.
This means we see the potential impact of delay without needing to caveat results regarding an assumed maximum possible sample size

\subsection{Efficiency metrics}

We define in the following sections several efficiency metrics that help us evaluate the performance of a SSR design in the presence of delay.

\subsubsection{Delay Impact}\label{Delay impact}

First, we define a performance metric that we refer to as the \textit{delay impact}.
This captures how likely a trial is to finish with a sample size greater than the re-estimated required sample size.
That is, the \textit{delay impact} is defined as the proportion of trials in the total number of performed simulations that conclude with $n_1 + n_\text{delay}$ as their final sample size.
This measure lies between 0 and 1 and a value closer to 1 would imply that it is highly likely that the number of pipeline participants, along with the first stage sample size, outnumbers the re-estimated required sample size.

\subsubsection{Root mean square error}

While the \textit{delay impact} can guide how delay impacts the final sample size, this does not completely capture how much efficiency the trial can lose as a result of delay. 
The goal of SSR can be thought of as attempting to estimate the true required sample size with precision.
That is, to have the final sample size be as close as possible to the true required sample size, $n_\text{oracle}$.
Therefore, the impact of delay may also be assessed by examining how the precision of the re-estimation process is affected, i.e., whether a delay in observing the primary outcome makes the final sample size drift apart from the oracle sample size, and if so, by how much. 
To quantify this, we compute the root mean square error (RMSE) of the final sample size.

The RMSE of the final sample size in the presence of delay can be defined as
\begin{equation*}
  RMSE = \sqrt{E(N^* - n_\text{oracle})^2},  
\end{equation*}
where $N^*$ is the random variable representing the final sample size obtained from Equation~(\ref{eq:ch4:N*}).
This is the square root of the MSE; as $MSE = Bias^2 + Variance$, it penalises designs that get the average re-estimated sample size incorrect and also ones that have higher variability in the re-estimated sample size.

Since the exact distribution of the final sample size accounting for delay is complex, we estimate the value of the RMSE by averaging the distance of the final sample size from $n_\text{oracle}$ across simulated trial replicates for our study.

\subsubsection{Cost}

Although RMSE is an effective measure at providing information on the accuracy of the re-estimated sample size accounting for delay, it fails to recognise that often an under-powered trial is considered to bear more serious consequences than an over-powered trial. 
Therefore, we also evaluate a metric that penalises an under-powered trial more than an over-powered trial for the same sample size difference. 

Let us define the \textit{Cost} to conducting a SSR design in the presence of a particular delay length as
$$ Cost = E\left[\frac{(N^* - n_\text{oracle})^2}{100 Power(N^*)}\right]. $$
Here, $Power(N^*)$ denotes the power of a two sample $t$-test with $N^*/2$ samples in each arm with the pre-specified $\delta$ and $\alpha$ values (3.5 and 0.05 respectively in our case study).
This metric can be viewed as a cost-benefit ratio, where, in this case, the cost is the loss of efficiency in terms of the distance from the ideal sample size for a given delay, and the benefit is the power of the trial. 
Note that by cost we do not refer to the specific financial cost associated with the treatment or the trial, rather the impact or cost of outcome delay on the trial.

A similar cost metric can be computed for a single-stage design.
However, in this case, the metric would take a constant value for particular error rate requirements.
This fixed value would depend on whether the planned sample size is higher or lower than $n_\text{oracle}$.

\subsection{Simulation study}

For our study of the impact of delay on a SSR design, we base our results on a simulation study, as the analytical distribution of the final sample size is complex.
For every simulation scenario, the initially planned sample size $n_{init}$, was computed assuming $\delta_0 = 3.5$ and $\sigma_{init}^2 = 10$.
All trials aimed to maintain 5\% significance level ($\alpha =  0.05$) and achieve 80\% power ($\beta = 0.2$) for a one-sided test.
This resulted in $n_{init} = 202$ participants in total in both arms as the initially planned sample size.
The recruitment rates were determined assuming a total recruitment period of 24 months for these $n_{init}$ samples, under both the considered uniform and linear recruitment patterns.

We primarily assume the interim analysis is planned after $n_1 = 70$ participants were recruited, following advice for external pilot trials \cite{Teare2014}.
However, an analysis of the sensitivity of the results to the interim analysis timing is also performed [See section \ref{sec 4.4}.
To explore the impact of delay, varying delay lengths were considered ranging from 0 to 24 months.

We then investigated three scenarios, given by
\begin{itemize}
    \item Case I:   $\sigma^2 = 8 < \sigma_{init}^2$.
    \item Case II: $\sigma^2 = 10 = \sigma_{init}^2$
    \item Case III:  $\sigma^2 = 12 > \sigma_{init}^2$.
\end{itemize}

In all cases, the true treatment effect was assumed to be $\delta = \delta_0 = 3.5$.

For each combination of parameter assumptions, 10,000 simulation replicates were run to obtain the distribution of the final sample size.
In each simulation, the first $n_1/2$ samples from each arm were drawn from $N(0, \sigma^2)$ and $N(\delta, \sigma^2)$ distributions for the control and treatment arm respectively.
The pooled sample variance was then computed, based on which the sample size was re-estimated.
The number of pipeline participants was then computed based on \textit{Computing the number of pipeline participants} section .
Next, Equation~(\ref{eq:ch4:N*}) was used to determine the final sample size incorporating delay.
Once the final sample size was determined, the rest of the stage 2 observations were simulated.
Finally, the test statistic was computed based on all $N^*$ outcome and the decision regarding whether to reject the null or not was made.

Note that, in the simulations we have not defined a maximum allowed sample size, meaning that there is no cap on the number of pipeline participants, the value of $N^*$, or the total recruitment period.
In practice, it is of course true that participant accrual is unlikely to go indefinitely; hence often SSR designs give a maximum allowed sample size in advance.
We have not fixed a maximum sample size in order to observe the full distribution of $N^*$ in the presence of delay, rather than truncating it to some maximum value.
This permits the results to be interpreted with the need to caveat regarding the assumed maximum allowed sample size.
The exact values of the RMSE and cost measures, as well as other parameters, obtained through simulations for selected delay lengths can be found in the Supplementary Materials. 
Results on the impact of delay on binary treatment outcomes are discussed in detail in the Supplementary Materials. 
The codes used for assessing the impact of delay on SSR can be found here: \url{https://github.com/AritraMukherjee/Sample-Size-Reestimation.git}

\section{Results}\label{SSR_delay impact}

\subsection{Impact of delay on the distribution of the final sample size}\label{SSR result}

\begin{figure}
\centering
\begin{subfigure}[]
  \centering
  \includegraphics[width=.4\linewidth]{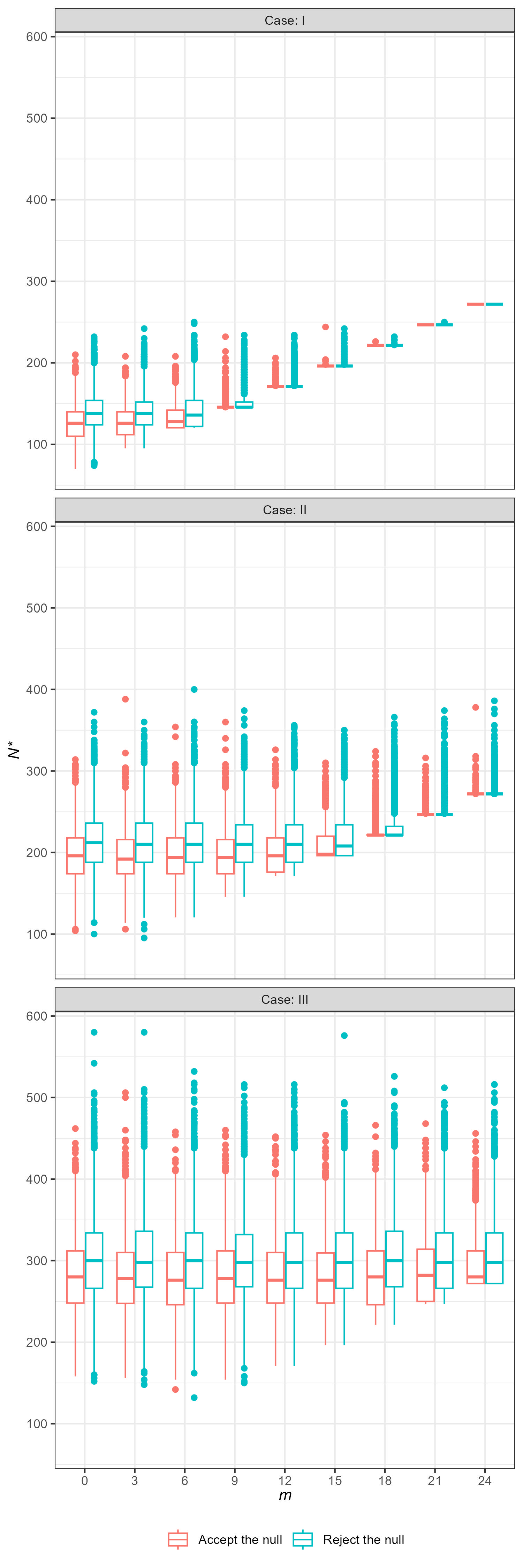}
   \label{fig1:N_SSR_uni}
\end{subfigure}
\begin{subfigure}[]
  \centering
  \includegraphics[width=.4\linewidth]{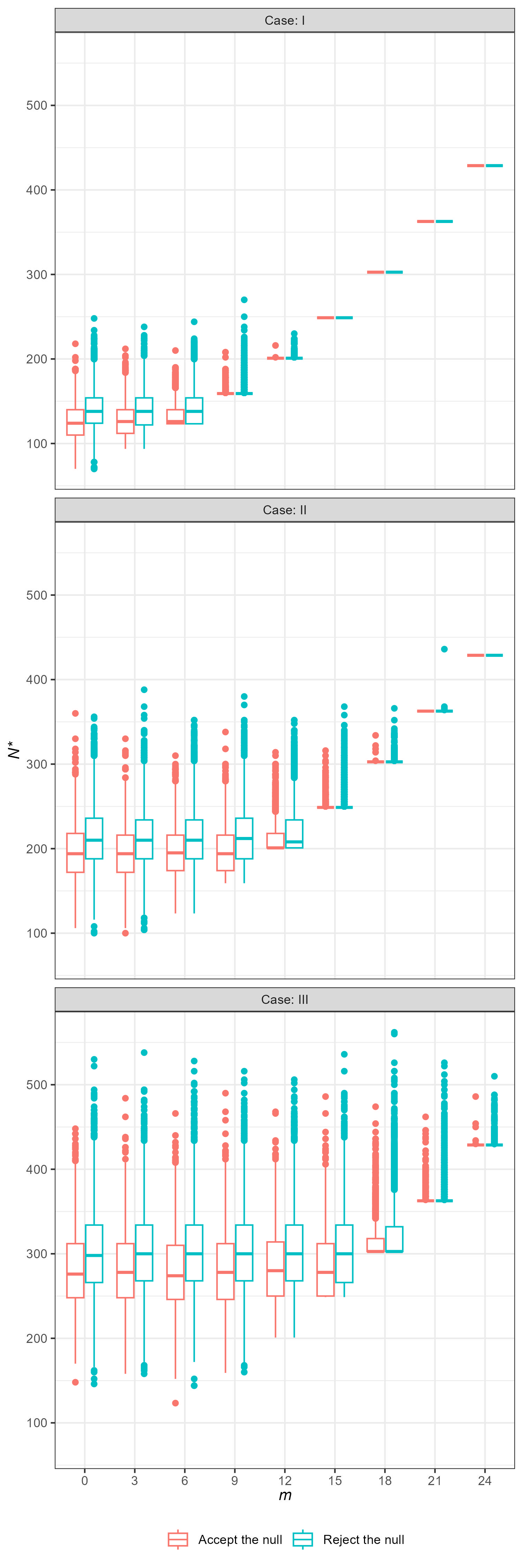}
   \label{fig2:N_SSR_lin}
\end{subfigure}
\caption{Distribution of the final sample size based on the decision to reject or accept the null under varying delay lengths for (a) uniformly and (b) linearly increasing  recruited samples and different values of $\sigma^2$.}
\label{fig1}
\end{figure}

First, we plot the distribution of the final sample size in the presence of delay for $m = 0, 3, 6, \dots, 24$. 
Figure~\ref{fig1} shows the distribution of $N^*$ following SSR for uniform and linear recruitment patterns respectively.
Since the underlying assumption in the simulation replicates treats the intervention as effective, the inability to detect such treatment effect can be costly.
A delayed outcome can add to this cost due to the pipeline participants. 
Hence we plot the impact of delay based on the decision regarding the null hypothesis.
In each figure, the blue boxplots denote the distribution of $N^*$ when the null hypothesis was rejected, and the red ones denote the same when the null was not rejected.

The primary finding evident from the results is that with an increase in $m$, the spread of the distribution of $N^*$ reduces, even if the median final sample size often remains similar. 
This is principally because with an increase in the delay length, the number of pipeline subjects increases considerably.
This increases the minimum attainable value for the sample size. 
By sub-case we observe the following
\begin{itemize}
    \item {\textbf{Case I}: $\sigma^2 = 8$}\\
    When $\sigma^2 = 8$, the trials are impacted heavily by delay.
    Here, the oracle design requires only 129 participants, compared to the initial sample size estimate of 202.
    Therefore, the re-estimated total sample size tends to be lower than the initially planned sample size.
    Further, especially in the presence of large delay, the total recruited samples at the end of the interim analysis will tend to surpass the re-estimated sample size.
    In other words, the chance is high that a larger number of participants are recruited than required; or alternatively, the final sample size would often be $n_1 + n_\text{delay}$ instead of $n_1 + n_{2*}$. 

    \item{\textbf{Case II}: $\sigma^2 = 10$}\\
    In trials where $\sigma^2 = 10$, it was observed that the spread of the sample size diminishes with the delay length. 
    However, the reduction is not as drastic as the case where $\sigma^2 = 8$. 
    The above statements are true for both uniform and linear recruitment patterns.
    However, the linear recruitment pattern leads to a larger number of pipeline subjects due to the increasing recruitment rate.
    
    \item{\textbf{Case III}: $\sigma^2 = 12$}\\
    Trials where the true population variance $\sigma^2 = 12$ are impacted the least by delay across the considered cases.
    Here, the oracle design requires a total of 290 participants.
    Therefore, for these trials, SSR tends to specify a re-estimated sample size larger than $n_{init}$.
    Consequently, the pipeline subjects are typically able to contribute positively to the final sample size. 
    We also observe that the spread of the sample size distribution remains relatively similar over varying delay lengths, with only very small changes observed for the higher delay values.

\end{itemize}

\begin{figure}[htbp]
    \centering
    \includegraphics[width=0.5\linewidth]{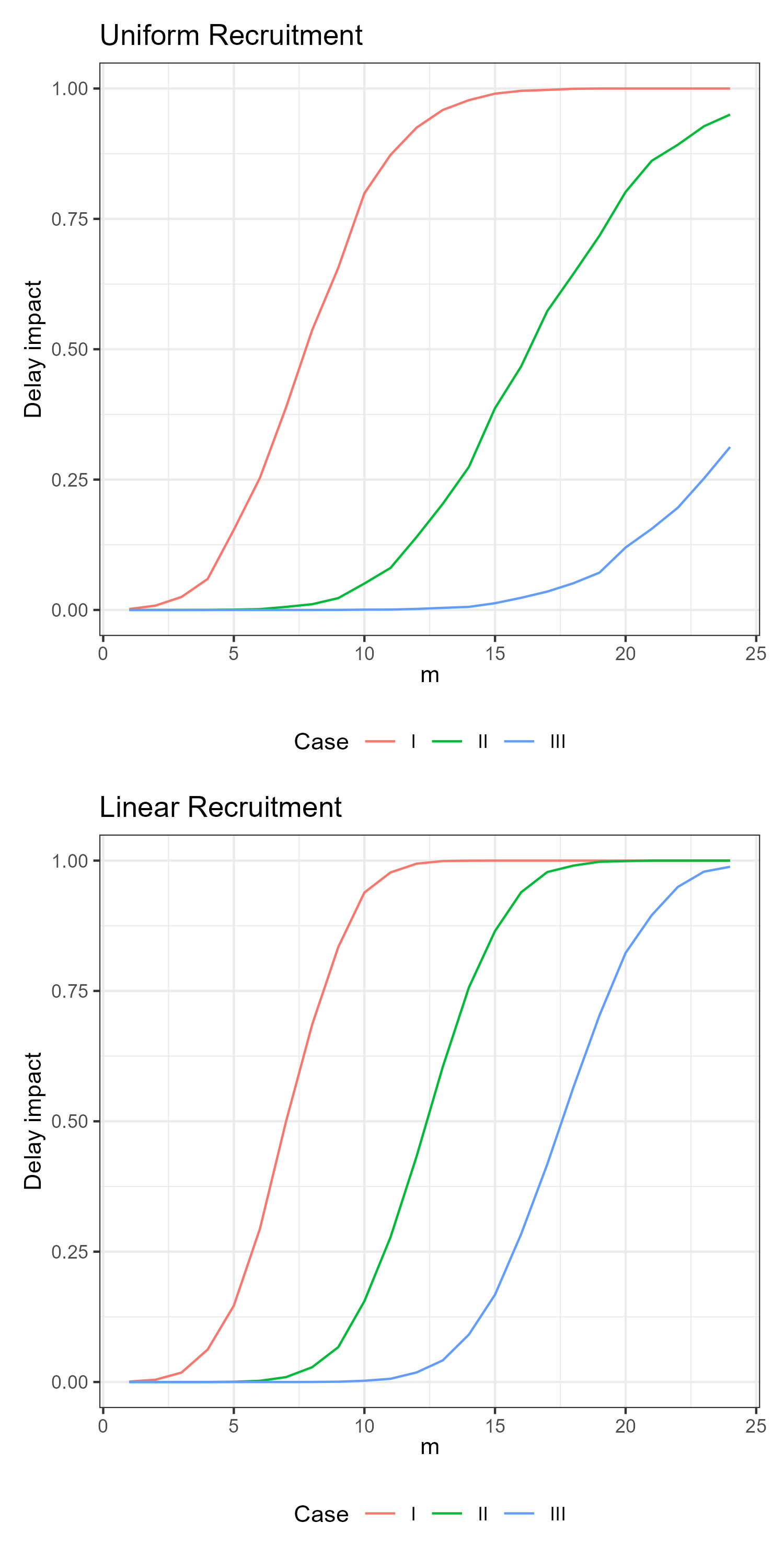}
    \caption{The `delay impact' for varying delay lengths $(m = 1, 2, \dots, 24)$ for $\sigma^2 = 8, 10, 12$, under uniform and linear recruitment patterns.}
    \label{fig:ch4:delay_impact_SSR}
\end{figure}

Building on the above, we plot the \textit{delay impact}, defined in \textit{Efficiency Metrics} section, in Figure~\ref{fig:ch4:delay_impact_SSR} for the aforementioned three cases.
It is evident from Figure~\ref{fig:ch4:delay_impact_SSR} that the \textit{delay impact} increases with $m$.
As an example, if we consider $m = 10$ months for Case I, the \textit{delay impact} is approximately 0.8 under uniform recruitment.
That is, there is an 80\% chance that the trial will finish with a sample size greater than that estimated as being required, resulting in a likely less efficient design.
This leads to an over-powered trial, adding to the financial burden of the trial.
It can be observed for trials where $\sigma^2 = 8$ or 10 the \textit{delay impact} is severe and quickly rises in $m$.
Whereas, for $\sigma^2 = 12$ the \textit{delay impact} is smaller, especially under uniform recruitment.

SSR under a linearly increasing recruitment pattern tends to suffer more from delay.
Here, due to higher pipeline participants, more trials are likely to end up with more than the estimated required number of participants, leading to an increased cost.
Even for $\sigma^2 = 12$, the \textit{delay impact} is observed to have a maximum of 99\%.
This reconfirms our previous inferences regarding the decreasing spread of the distribution of $N^*$.

\subsection{Impact of delay on the root mean square error}



\begin{figure*}[htbp]
     \centering
     \includegraphics[width=\textwidth]{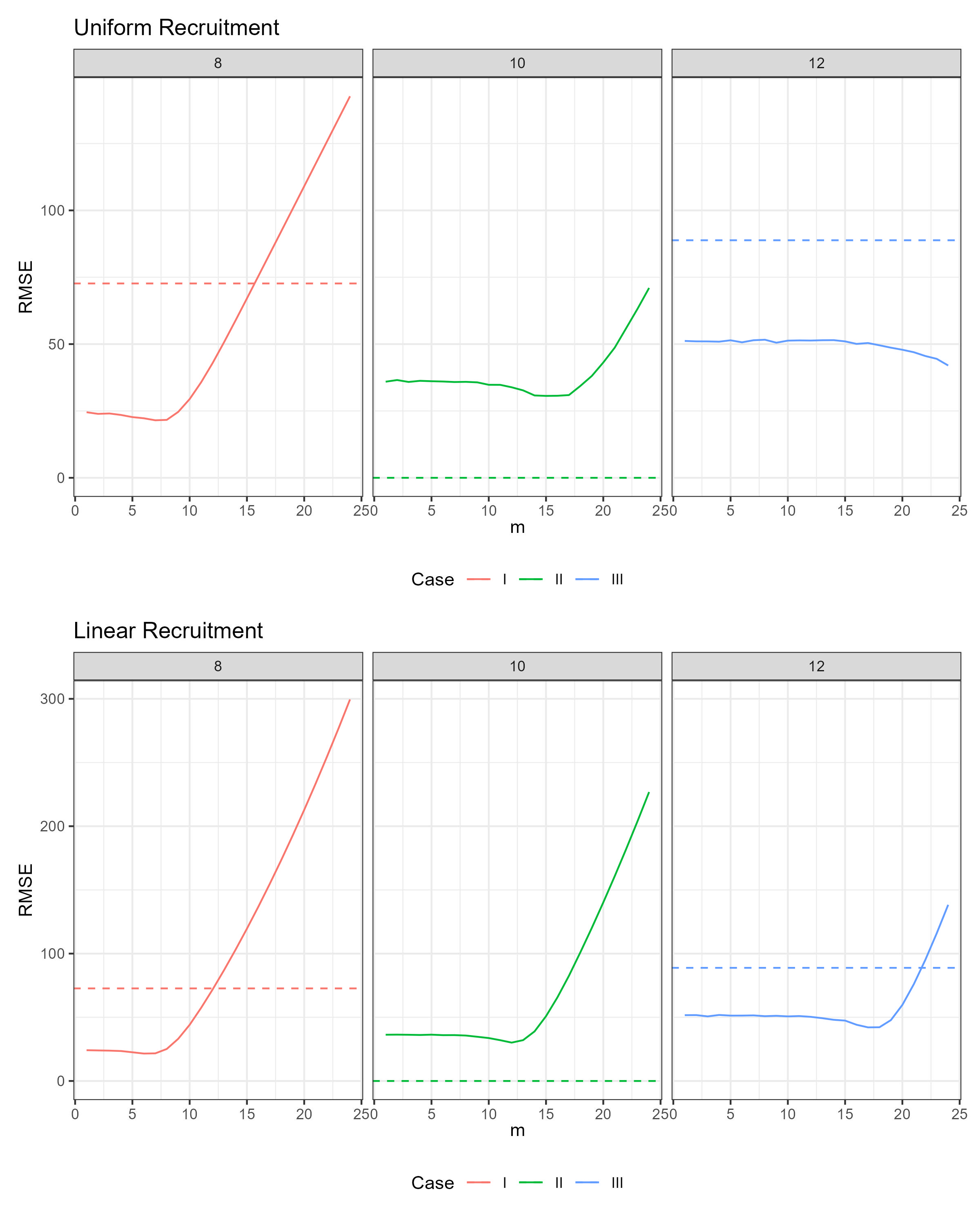}
        \caption{RMSE for varying delay lengths $(m = 1, 2, \dots, 24)$ for $\sigma^2 = 8, 10, 12$, under uniform and linear recruitment patterns.
        The dotted line in each graph represents the RMSE for a single-stage design.}
        \label{fig:ch4:MSE_SSR}
\end{figure*}

Figure~\ref{fig:ch4:MSE_SSR} plots the RMSE for Cases I-III for $m = 1, 2, \dots, 24$ based on 10,000 simulations for each parameter combination specified in \textit{Simulation Study} section. 
The dotted line in each graph represents the RMSE for a single-stage design, which reduces to just the difference between the single-stage sample size and the oracle sample size and is thus constant across delay lengths for given $\alpha$, $\beta$, $\sigma_{init}$, and $\delta$ values.

It can be observed from Figure~\ref{fig:ch4:MSE_SSR} that for smaller delay lengths the RMSE remains relatively constant (with variations attributable to sampling variation).
However, with increase in the delay length, the RMSE increases.
For Cases I and II it increases rapidly for $m$ greater than 15 months. 
This reflects the drift of $N^*$ from $n_{oracle}$ for the trial and therefore losing much of the efficiency of the design in the process.
If the recruitment is assumed to be linearly increasing, then the impact is even greater, with a rapid increase in the RMSE happening sooner at approximately $m = 12$ months. 

When $\sigma^2 = 8$, the RMSE observes a sharp increase beyond a 9-month delay period. 
A similar pattern can be observed for $\sigma^2 = 10$ where, the increase in RMSE starts after a 18-months delay.
It can be inferred that in this case, for a large delay length (e.g., $m > 9$ months), the final sample size usually does not correctly represent the true sample size required by the trial, leading to an over-powered trial on average.

\subsection{Impact of delay on the cost}\label{SSR cost metric}

\begin{figure*}[htbp]
     \centering
     \includegraphics[width=\textwidth]{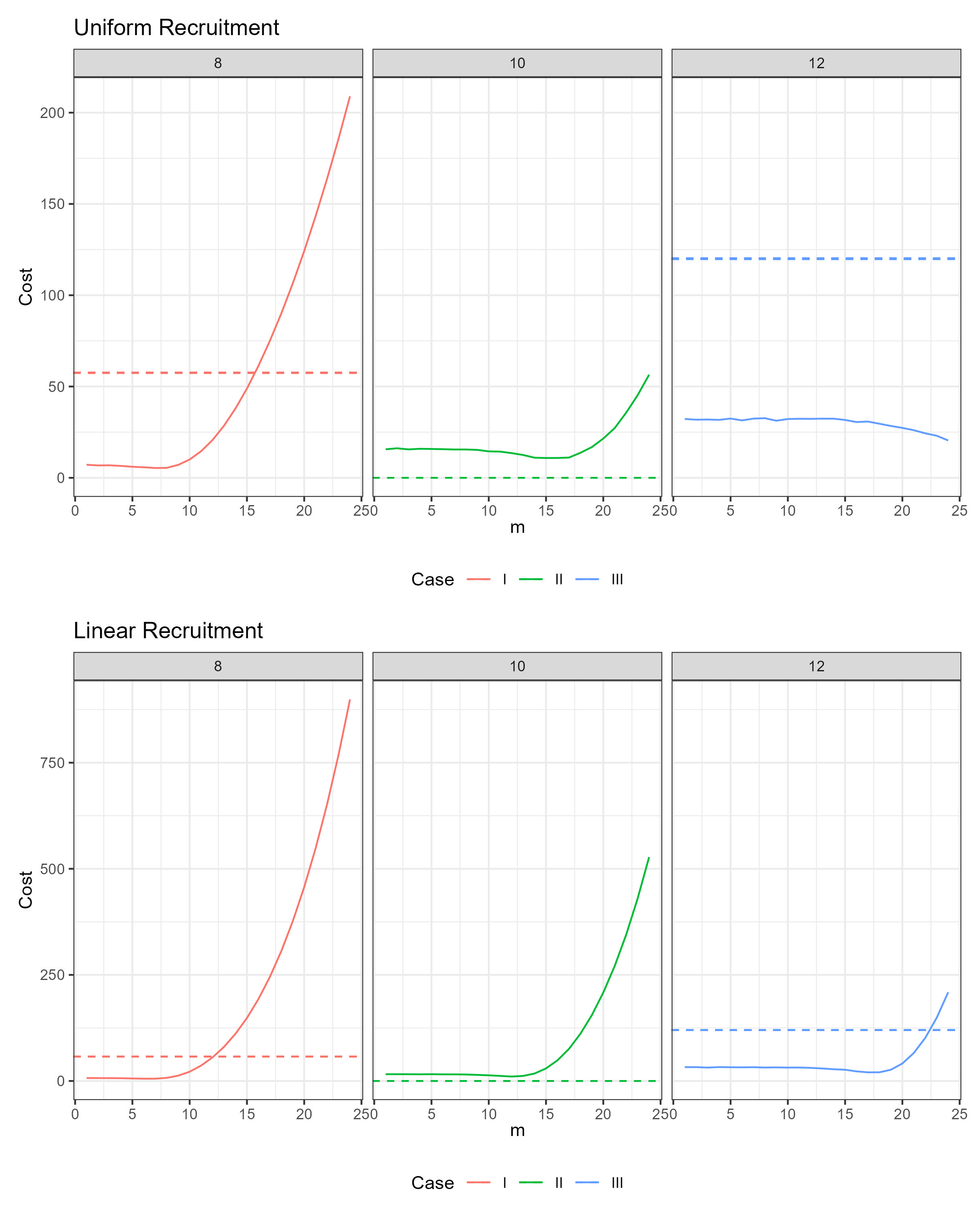}
        \caption{The `Cost' for varying delay lengths $(m = 1, 2, \dots, 24)$, for $\sigma^2 = 8, 10, 12$, under uniform and linear recruitment patterns.}
        \label{fig:ch4:Cost_SSR}
\end{figure*}

In this section, we plot the cost metric in Figure~\ref{fig:ch4:Cost_SSR} for both single-stage and blinded SSR designs.
The figure shows that there exists almost an exponential increase in the cost for greater delay lengths. 
This figure looks very similar to the plot of RMSE as shown in Figure~\ref{fig:ch4:MSE_SSR}, reconfirming our previous observations.

The impact of delay in terms of this cost metric is highly dependent on $\sigma^2$. 
When $\sigma^2 = 8$, a higher cost is suffered in the presence of large delay ($m > 15$ months, for Cases I and II), aligning with the previous inferences.
The only case when SSR is comparatively beneficial compared to a single-stage design, irrespective of the considered delay lengths, is when $\sigma^2 = 12$.
In fact, in this case, under uniform recruitment observing a delay for the primary outcome may add to the efficiency of the trial compared to the alternative option of pausing recruitment to conduct the interim analysis, as it can lead to recruitment of a number of participants closer to the oracle sample size while awaiting treatment outcome data. 

In addition, it can be seen that the recruitment pattern does influence the efficiency losses.
In general, under linear recruitment larger costs are suffered as a greater number of pipeline subjects are usually present.

Here, one interesting point to note in Figures~\ref{fig:ch4:MSE_SSR} and \ref{fig:ch4:Cost_SSR} is that there is a small dip in the RMSE and cost values before the curves start to rise rapidly.
For example for Case II, the value of RMSE falls between $10 \le m \le 18$ and attains a minimum at $m = 15$ months.
If we take a closer look at the tables in the Supplementary Materials, it can be seen that for $m = 15$ months, $n_\text{delay} = 126$. 
Along with the 70 first stage participants, the final sample size then results in 196 participants, which is very close to the required oracle sample size of approximately 202.
Hence, the minimum value of the final sample size increases to 196.
Thus, in this case, the variability along with the RMSE value reduces as compared to the RMSE for $m = 0$ months.
Thus, we see a small dip in the values of RMSE as well as the cost in that region of $m$ values.

\subsection{Effect of delay for different first stage sample sizes}\label{sec 4.4}

\begin{figure}
\centering
\begin{subfigure}[]
  \centering
  \includegraphics[width=.4\linewidth]{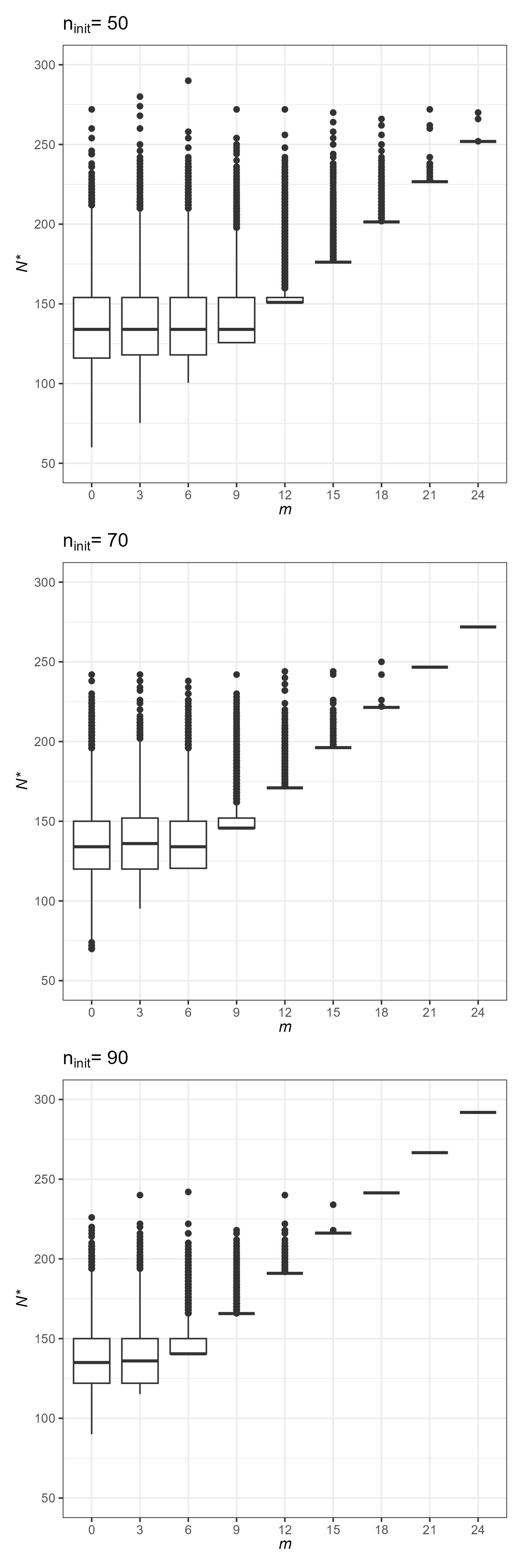}
   \label{fig:Ch4:N_n1_overestimated_sigma}
\end{subfigure}
\begin{subfigure}[]
  \centering
  \includegraphics[width=.4\linewidth]{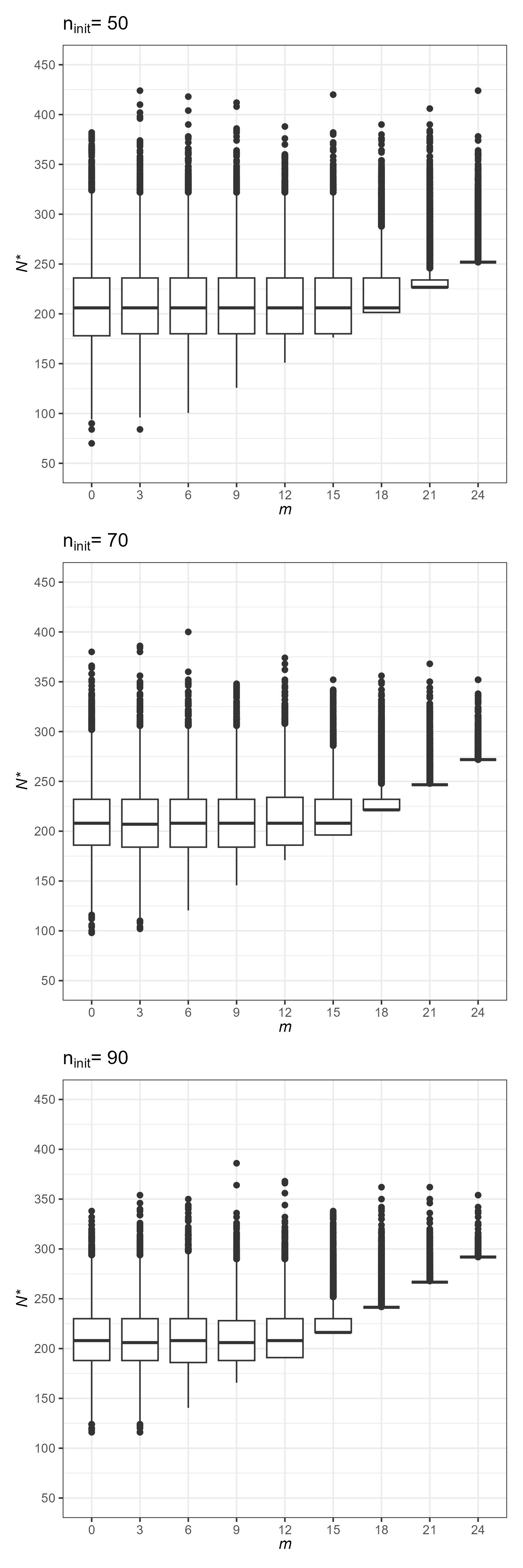}
   \label{fig:Ch4:N_n1_correct_sigma}
\end{subfigure}
\caption{Final sample sizes for varying delay lengths for different first stage sample sizes ($n_1 = 50, 70, 90$) assuming uniform recruitment when (a) $\sigma^2 = 8$ and (b) $\sigma^2 = 10$ respectively. The initially planned sample size was 101 per arm for both the cases, whereas the oracle sample size in this case is 65 per arm for case (a) and 101 for case (b). }
\label{fig5}
\end{figure}

The results so far showed that the impact of delay is highly sensitive to the value of the nuisance parameter, as this helps determine the initially planned sample size (and thus influence the recruitment rate as per our model assumptions).
In order to obtain a reliable estimate of the nuisance parameter, the sample size at the re-estimation point needs to be chosen carefully. 
For external pilot trials, Teare et al. \cite{Teare2014} suggested using 35 samples in each arm to estimate $\sigma^2$ in the normal outcome case (for standardised treatment effect sizes of 0.2, 0.35 or 0.5 and 90\% power).
Although our simulation assumption suggests that for the power requirement of 80\% and our assumed effect size, we might need smaller first stage sample sizes, in order to account for variability in estimation of the pooled SD, we started with the initial internal pilot size of 70.
In this section, we seek to observe how varying the first stage sample size impacts the final sample size in the presence of delay.
We plot the final sample size $N^*$ when $\sigma^2 = 8$ or 10, the two previously considered cases more heavily impacted by delay.
Figure~\ref{fig5} plots the final sample sizes for varying delay lengths for different first stage sample sizes, specifically $n_1 = 50, 70, 90$.
We plot the results for the case of a uniform recruitment pattern. 

It can be observed that, as expected, the variation in the final sample sizes reduces as the first stage sample size $n_1$ increases.
Also, the minimum value for the final sample size rises quickly for a higher $n_1$.
This is mainly because a higher $n_1$ value along with the number of pipeline participants quickly increases the minimum attainable final sample size with increasing delay.
The impact is more severe when $\sigma^2 = 8$, however, similar trends can be seen when $\sigma^2 = 10$.

If the recruitment becomes linearly increasing, it can be said that the impact will be more severe due to the increasing recruitment pattern yielding a higher number of pipeline participants.
Further a higher $n_1$ ensures a higher recruitment rate at that time for a linear recruitment as compared to a uniform recruitment pattern.

In order to reach a balance between the severity of delay impact and higher variability in the sample size, our results arguably coincide with the findings of Teare et al.
That is, 35 samples in each arm appears to strike an appropriate balance between obtaining a reasonably reliable estimate for $\sigma^2$ as well as limiting the impact of delay.

\section{Discussion}

Adaptive designs have gained much popularity in recent years due to the potential trial efficiency advantages they can provide.
However, most of the adaptive design literature assumed that the primary treatment outcome is available immediately post recruitment.
In practice, it may take some time to observe the primary treatment outcome.
Such outcome delays can result in much of the expected efficiency gains of an adaptive design being lost.
Earlier studies indicate for group-sequential designs that a delay which is more than 50\% of the total recruitment length may result in an inefficient adaptive trial, with all of the added efficiency lost due to delay.

SSR can be a powerful tool to ensure a trial meets its desired power requirement.
However, a delayed outcome has the potential to result in an over-powered trial, adding to the financial burden of the trial.
While methods exist to incorporate short-term information in re-estimating the sample size, they are not predominantly used in practice.
In this paper, we sought to observe the impact of delayed outcomes on the efficiencies that SSR provides.
We have considered both continuous and binary outcome variables to demonstrate this impact. 
We assessed the efficiency lost due to delayed outcomes through the distribution of the final sample size accounting for delay and the RMSE of the final sample size.
We introduced a cost metric that penalises an under-powered trial as a result of SSR more than an over-powered trial, in order to arguably better capture the true extent of efficiency loss experienced by a trial.
Furthermore, we also have defined the \textit{delay impact} to be able to determine the likelihood of a trial ending up with more participants being recruited in the trial than required for a particular delay length.

The results show that outcome delay does impact the efficiency of SSR, which if ignored at the planning stage can cause the trial heavy financial disadvantages as well as waste of resources.
However, this efficiency loss is heavily dependent on the initial sample size specified.
When the true variance parameter is smaller than initially specified, the reduced required sample size makes the trial more susceptible to delay.
By contrast, if the variance is large, the large required sample size makes delay less impactful.
In general, it can be said that if there is a chance that the initial sample size is overestimated, an SSR design can be costly in presence of an outcome delay.
Therefore, in presence of delay, starting with a relatively smaller sample size for the interim analysis might be a better approach.

The uniform recruitment assumed to estimate the number of pipeline participants can be representative of small scale single centered trials.
However, in multi-centered trials the recruitment rate is likely to increase gradually as new centres open up and reaches a constant uniform rate as all the centres become fully operational.
As an extreme case of the aforementioned recruitment pattern, we assumed a linearly increasing recruitment pattern, where the recruitment rate does not plateau.

It is to be noted that the results in this paper may arguably be viewed as reflecting only small changes in the underlying parameter values.
Since the impact of delay was still highly sensitive to these parameter values, higher fluctuations could clearly gravely impact the trial efficiency.
Furthermore, the results in this paper are based on the assumption that there is no cap on the recruitment, which does not reflects reality.
In this paper, we observed the maximum loss possible for a given delay length for specific parameter values used in the sample size calculation.
However, note that, this loss can be restricted, especially in Case I, by capping the maximum number of recruitments possible in the trial.
But, this efficiency loss will be sensitive to the maximum sample size specified.

We recognise that, current guidelines (e.g. EMA reflection paper on adaptive designs\cite{EMAguideline}) suggest under-powered trials are a greater problem than over powered trials.
However, the goal in this study is to achieve the desired power as accurately as possible through reestimation.
Therefore, the overpowered study has also been considered undesirable as an underpowered study.
However, the cost proposed in the study naturally penalises underpowered study more severely as compared to an overpowered study, providing a better picture to the true extent of loss possible.

Our results clearly indicate that precise estimates of the nuisance parameter can largely influence how delay might impact the efficiency of a trial.
Hence, it becomes crucial to be able to efficiently determine the nuisance parameter as early as possible in order to minimise damage.
Teare et al. \cite{Teare2014} suggested 35 participants per arm to be a reasonable sample size to provide an accurate estimate of the variance of a normally distributed outcome (for standardised treatment effect sizes of 0.2, 0.35 or 0.5 and a 90\% power).
Therefore, our simulations for continuous outcome variables assumes a first stage sample size of 35 per arm in the beginning.
However, we observed the impact of delay on the choice of the first stage sample size (or the timing of the interim analysis) for continuous outcomes.
The results given here extend the findings of Teare et al. in an interesting manner; as expected, increasing the first stage sample size naturally still translated into a lower variability in the re-estimated sample size when considering delay.
However, the impact of delay increased as a function of the first stage sample size as it has a potential of ending up with a much larger sample size than required.
Importantly, it could well be argued based on the given results that in order to strike a balance between these two conflicting interests, conducting the interim after recruiting 35 participants in each arm remains a reasonable choice.
Note that the work by Teare et al. is set in an external pilot trial setting, rather than an internal pilot for blinded SSR.
Thus, the results discussed here extend the work of Teare et al., confirming that an internal pilot sample size of 35 participants per arm is applicable for blinded SSR as well.

In conclusion, a delay in observing the treatment outcome can impact a SSR design's efficiency negatively.
However, this is highly dependent on the recruitment rate and initially planned sample size.
Carefully planned simulation studies are necessary in planning such trial to truly observe the extent of harm that a delayed outcome can do to the trial.
In order to minimise the impact, 35 participants in each arm for continuous outcome data seems to be a reasonable choice for the interim.
It is also important to note that the impact of delay on SSR is lower than designs which allow early stopping such as Simon's two-stage design, a two-arm or multi-arm group-sequential design.

\section*{Conflict of interest}
The authors declare that there is no conflict of interest.

\section*{Data availability}
The study uses simulated datasets for results. The codes which support the findings of this study are openly available in Github at \url{https://github.com/AritraMukherjee/Sample-Size-Reestimation}.

\section*{Ethics approval}
This research did not require any ethical approval required as the study did not involve any human or animal subjects.

\section*{Funding acknowledgement}

JMSW and AM is funded by a NIHR Research Professorship (NIHR301614).


\end{document}


\maketitle

\onehalfspacing

\section{Impact of delay on SSR design with continuous treatment outcomes}

This section contains the exact values for \textit{delay impact}, $MSE$, the cost metric and empirical power values for different $m$ values (in particular, $m=0,3,6,9,\dots, 24$). 
Table \ref{tab:SSR_uni} provides these values along with the oracle sample size and average final sample sizes from the 10,000 simulation under a uniform recruitment assumption, whereas, table \ref{tab:SSR_lin} provides the same for a linearly increasing recruitment pattern.

It can be seen from tables \ref{tab:SSR_uni}-\ref{tab:SSR_lin} that, the average re-estimated sample size increases with delay.
This is because, as the \textit{delay impact} increases a greater proportion of trials ends up with $n_1+n_{delay}$ sample.
This gives a rise to the minimum attainable re-estimated sample size thus increasing the average re-estimated sample size. 

For trials in which $\sigma_{init}^2>\sigma^2$, they tend to become overpowered quickly due to the increase in the sample size. 
The power can increase to as big as 97\% from 80\% for a sufficiently large delay (24 months). 
The power increases from 80 to 86\% for a trial with correctly specified $\sigma_{init}^2=\sigma^2$ over increasing delay lengths. 
The power remains relatively constant for the third scenario. 
However, for linear recruitment, the power is further influenced due to the increase in pipeline subjects. 

\begin{landscape}
\begin{longtable}[c]{@{}ccccccccccccc@{}}
\caption{Impact of delay $(m)$ on Efficiency and cost parameters for a SSR design with Uniform recruitment.
Here, all the 10,000 simulated designs for every $m$ maintain $\alpha=0.05, \beta=0.2, \delta_0=\tau=3.5$.
The initial sample size is computed based on $\sigma_0^2=10$ as 202 and the interim is conducted after 70 patients are recruited across both arms.}
\label{tab:SSR_uni}\\
\toprule
\textbf{$\sigma^2$} & \textbf{\begin{tabular}[c]{@{}c@{}}Empirical\\ Power\end{tabular}} & \textbf{\begin{tabular}[c]{@{}c@{}}$m$\end{tabular}} & \textbf{$n_{oracle}$} & \textbf{\begin{tabular}[c]{@{}c@{}}Average \\ $N^*$\end{tabular}} & \textbf{$n_{delay}$} & \textbf{\begin{tabular}[c]{@{}c@{}}Average\\ $\sigma_*^{2}$\end{tabular}} & \textbf{$MSE_{single}$} & \textbf{$MSE_{SSR}$} & \textbf{$Cost_{single}$} & \textbf{$Cost_{SSR}$} & \textbf{\begin{tabular}[c]{@{}c@{}}Delay \\ impact\end{tabular}} \\ \hline
\endhead
\bottomrule
\endfoot
\endlastfoot

8 & 0.7995 & 0  & 129.2 & 136.75 & 0 & 8.17 & 5281.89 & 579.98 & 57.57 & 6.89 & 0 \\
 & 0.8022 & 3    &   & 136.73 & 25.23 & 8.17 &  & 580.08 &  & 6.86 & 0.02 \\
 & 0.8183 & 6    &   & 139.81 & 50.47 & 8.17 &  & 483.99 &  & 5.64 & 0.26 \\
 & 0.8427 & 9    &   & 151.11 & 75.70 & 8.17 &  & 606.45 &  & 7.06 & 0.66 \\
 & 0.8863 & 12    &   & 171.94 & 100.94 & 8.17 &  & 1847.38 &  & 20.82 & 0.92 \\
 & 0.9228 & 15    &   & 196.24 & 126.17 & 8.17 &  & 4498.53 &  & 48.88 & 0.99 \\
 & 0.9399 & 18    &   & 221.40 & 151.41 & 8.16 &  & 8502.18 &  & 89.99 & 1 \\
 & 0.9645 & 21    &   & 246.57 & 176.64 & 8.18 &  & 13788.43 &  & 143.33 & 1 \\
 & 0.9712 & 24    &   & 271.78 & 201.88   & 8.16 &  & 20348.18 &  & 208.87 & 1 \\* \midrule
10 & 0.7927 & 0   & 201.88 & 208.79 & 0 & 10.11 & 0 & 1291.35 & 0 & 15.63 & 0 \\
 & 0.8037 & 3   &    & 208.89 & 25.23 & 10.11 &  & 1301.79 &  & 15.79 & 0 \\
 & 0.8012 & 6   &    & 209.78 & 50.47 & 10.13 &  & 1344.43 &  & 16.23 & 0 \\
 & 0.8025 & 9   &    & 209.85 & 75.70 & 10.13 &  & 1290.77 &  & 15.48 & 0.02 \\
 & 0.7934 & 12   &    & 210.57 & 100.94 & 10.10 &  & 1115.52 &  & 13.18 & 0.14 \\
 & 0.8106 & 15   &    & 216.98 & 126.17 & 10.11 &  & 907.22 &  & 10.48 & 0.39 \\
 & 0.8357 & 18   &    & 230.56 & 151.41 & 10.12 &  & 1145.08 &  & 13.30 & 0.65 \\
 & 0.8682 & 21   &    & 249.65 & 176.64 & 10.11 &  & 2384.59 &  & 27.40 & 0.86 \\
 & 0.8923 & 24   &    & 272.74 & 201.88 & 10.12 &  & 5050.42 &  & 56.61 & 0.95 \\* \midrule
12 & 0.7989 & 0   & 290.71 & 298.55 & 0 & 12.10 & 7890.23 & 2642.23 & 120.13 & 32.41 & 0 \\
 & 0.7978 & 3   &    & 297.21 & 25.23 & 12.07 &  & 2619.70 &  & 32.15 & 0 \\
 & 0.7988 & 6   &    & 298.88 & 50.47 & 12.10 &    & 2640.92 &  & 32.23 & 0 \\
 & 0.7853 & 9   &    & 297.53 & 75.70 & 12.08 &    & 2572.66 &  & 31.57 & 0 \\
 & 0.7987 & 12   &    & 298.68 & 100.94 & 12.10 &    & 2583.37 &  & 31.61 & 0 \\
 & 0.7992 & 15   &    & 298.13 & 126.17 & 12.09 &    & 2569.34 &  & 31.27 & 0.01 \\
 & 0.8023 & 18   &    & 298.56 & 151.41 & 12.08 &    & 2457.84 &  & 29.67 & 0.05 \\
 & 0.8045 & 21   &    & 301.35 & 176.64 & 12.09 &    & 2205.63 &  & 26.10 & 0.16 \\
 & 0.812 & 24   &    & 306.45 & 201.88 & 12.06 &    & 1824.05 &  & 21.09 & 0.33 \\* \bottomrule
\end{longtable}
\end{landscape}

\begin{landscape}
\begin{longtable}[c]{@{}ccccccccccccc@{}}
\caption{Impact of delay $(m)$ on Efficiency and cost parameters for a SSR design with Linear recruitment
Here, all the 10,000 simulated designs for every $m$ maintain $\alpha=0.05, \beta=0.2, \delta_0=\tau=3.5$.
The initial sample size is computed based on $\sigma_0^2=10$ as 202 and the interim is conducted after 70 patients are recruited across both arms.}
\label{tab:SSR_lin}\\
\toprule
\hline
\textbf{$\sigma^2$} & \textbf{\begin{tabular}[c]{@{}c@{}}Empirical\\ Power\end{tabular}} & \textbf{\begin{tabular}[c]{@{}c@{}}$m$\end{tabular}} & \textbf{$n_{oracle}$} & \textbf{\begin{tabular}[c]{@{}c@{}}Average \\ $N^*$\end{tabular}} & \textbf{$n_{delay}$} & \textbf{\begin{tabular}[c]{@{}c@{}}Average\\ $\sigma_*^2$\end{tabular}} & \textbf{$MSE_{single}$} & \textbf{$MSE_{SSR}$} & \textbf{$Cost_{single}$} & \textbf{$Cost_{SSR}$} & \textbf{\begin{tabular}[c]{@{}c@{}}Delay \\ impact\end{tabular}} \\ \hline
\endhead

\bottomrule
\endfoot
\endlastfoot

8 & 0.8014 & 0 & 129.20 & 136.38 & 0.00 & 8.16 & 5281.89 & 587.34 & 57.53 & 6.98 & 0.00 \\
 & 0.7961 & 3  &   & 136.65 & 23.64 & 8.16 &    & 571.35 &  & 6.77 & 0.02 \\
 & 0.8141 & 6  &   & 140.32 & 53.34 & 8.16 &    & 459.85 &   & 5.36 & 0.30 \\
 & 0.8651 & 9  &   & 161.31 & 89.10 & 8.15 &    & 1084.90 &   & 12.45 & 0.84 \\
 & 0.9185 & 12  &   & 200.96 & 130.91 & 8.16 &    & 5154.42 &   & 55.67 & 0.99 \\
 & 0.9614 & 15  &   & 248.69 & 178.78 & 8.17 &    & 14292.89 &   & 148.37 & 1.00 \\
 & 0.9827 & 18  &   & 302.68 & 232.70 & 8.15 &    & 30099.67 &   & 305.90 & 1.00 \\
 & 0.9942 & 21  &   & 362.57 & 292.68 & 8.17 &    & 54493.17 &  & 548.36 & 1.00 \\
 & 0.9982 & 24  &   & 428.58 & 358.72 & 8.15 &    & 89677.13 &   & 898.70 & 1.00 \\* \midrule
10 & 0.7934 & 0  & 201.88 & 209.26 & 0.00 & 10.12 &  0 & 1320.92 &  0 & 15.97 & 0.00 \\
 & 0.7918 & 3  &   & 208.67 & 23.64 & 10.11 &    & 1299.98 &    & 15.76 & 0.00 \\
 & 0.7935 & 6  &   & 209.12 & 53.34 & 10.12 &    & 1311.21 &    & 15.87 & 0.00 \\
 & 0.8008 & 9  &   & 209.59 & 89.10 & 10.11 &    & 1242.04 &    & 14.81 & 0.07 \\
 & 0.8251 & 12  &   & 219.26 & 130.91 & 10.12 &    & 896.49 &    & 10.35 & 0.42 \\
 & 0.8672 & 15  &   & 251.53 & 178.78 & 10.11 &    & 2561.55 &    & 29.38 & 0.87 \\
 & 0.9161 & 18  &   & 302.83 & 232.70 & 10.12 &    & 10194.49 &    & 111.00 & 0.99 \\
 & 0.9534 & 21  &   & 362.69 & 292.68 & 10.10 &    & 25859.58 &    & 271.18 & 1.00 \\
 & 0.9769 & 24  &   & 428.72 & 358.72 & 10.12 &    & 51457.47 &    & 527.50 & 1.00 \\* \midrule
12 & 0.789 & 0  & 290.71 & 298.46 & 0.00 & 12.10 & 7890.23 & 2587.31 & 120.12 & 31.71 & 0.00 \\
 & 0.7871 & 3  &    & 297.46 & 23.64 & 12.07 &   & 2552.79 &  & 31.33 & 0.00 \\
 & 0.8005 & 6  &    & 297.53 & 53.34 & 12.08 &   & 2622.84 &  & 32.13 & 0.00 \\
 & 0.7932 & 9  &    & 297.42 & 89.10 & 12.07 &   & 2579.55 &  & 31.71 & 0.00 \\
 & 0.7876 & 12  &    & 297.46 & 130.91 & 12.07 &   & 2547.19 &  & 31.13 & 0.02 \\
 & 0.8000 & 15  &    & 301.64 & 178.78 & 12.08 &   & 2168.67 &  & 25.62 & 0.17 \\
 & 0.8237 & 18  &    & 320.45 & 232.70 & 12.08 &   & 1762.22 &  & 20.34 & 0.57 \\
 & 0.8731 & 21  &    & 365.67 & 292.68 & 12.09 &   & 5767.14 &  & 65.91 & 0.90 \\
 & 0.9144 & 24  &    & 428.92 & 358.72 & 12.06 &   & 19111.18 &  & 208.94 & 0.99 \\* \bottomrule
\end{longtable}
\end{landscape}

\section{Impact of delay on SSR design with binary treatment outcomes}

\subsection{SSR for binary treatment outcome}

The paper describes in detail SSR for continuous outcome variables.
For a binary response variable, the required sample size depends not only on the specified values of the type I error rate, power, and clinically relevant difference and the nuisance parameter, but also on the precise underlying success probabilities in the two arms.
Friede and Kieser \cite{Friede2004} proposed a blinded SSR process to re-estimate the sample size in this scenario. 
We present the effect of delay on such designs in this section.

Specifically, let us consider a clinical trial comparing two treatments based on a binary outcome variable. 
The success rates in each group are denoted as $\pi_1$ and $\pi_2$ respectively. 
Suppose there are to be $n/2$ samples observed in each treatment arm such that the total sample size for the trial is $n$.
$X_i$, the number of successes in group $i = 1, 2$, is then binomially distributed with parameters $n/2$ and $\pi_i$.
The parameter of interest is the absolute difference in the success probabilities, $\delta = \pi_2 - \pi_1$, and the hypotheses under test at level $\alpha$ and power $(1-\beta)$ for $\delta=\delta_1$ are assumed to be $H_0 : \delta \le 0$ versus $H_0 : \delta > 0$.
A normal approximation test can be used to test the above hypotheses, with the test statistic 
\begin{equation*}
    U = \sqrt{\frac{1}{2}*\frac{n}{2}} \frac{\hat{\pi}_2 - \hat{\pi}_1}{\sqrt{\bar{\pi}(1-\bar{\pi})}},
\end{equation*}
where $\hat{\pi}_i=\frac{X_i}{n/2}$ are the observed proportions of successes in the arms and $\overline{\pi}$ is the observed pooled success rate across arms, computed as $\frac{X_1 + X_2}{n}$.
We reject the null at significance level $\alpha$ if $U > \Phi^{-1}(1 - \alpha)$ \cite{KieserM}.

The sample size required for the above test for a significance level of $\alpha$ and power of $1 - \beta$ is typically computed as
\begin{equation}\label{eq: sample_size_binary}
    n= 2*\frac{[\Phi^{-1}(1-\alpha)\sqrt{2 \bar{p}(1-\bar{p})} + \Phi^{-1} (1-\beta)\sqrt{p_1(1 - p_1) + p_2(1 - p_2)}]^2}{(p_2 - p_1)^2},
\end{equation}
where $p_1$ and $p_2$ are pre-specified estimates of $\pi_1$ and $\pi_2$ and $\bar{p} = (p_1 + p_2)/2$ \cite{KieserM}.

As before, if the true values of $\pi_1$ and $\pi_2$ were known then the oracle sample size derived from the above formula, $n_\text{oracle}$, can be readily calculated.
However, at the planning stage the values of the $\pi_i$ are unknown and the $p_i$ used in the calculation may be subject to substantial uncertainty.

Let us assume that the initial sample size estimate for the trial is $n_{init}$ based on equation \ref{eq: sample_size_binary}.
After $n_1$ patients have been recruited in both the control and the treatment arm, we estimate the value of the pooled success rate and re-estimate the sample size based on this.
Usually, $n_1$ is considered to be a fraction of $n_{init}$ or pre-specified at the design stage.
By estimating only a pooled success rate the blinding of the treatment allocations can remain intact.
Here, the pooled success rate can be estimated as $p = (X_{11}+X_{12})/n_1$ , where, $X_{11}$ and $X_{12}$ denotes the total number of successes in the first stage, and the re-estimated sample size ($n_*=n_1+n_{2*}$) is given as
\begin{equation}\label{N*}
   n_* = 2*\frac{\{\Phi^{-1} (1-\alpha) + \Phi^{-1} (1-\beta)\}^2}{\delta_1^2}2p(1-p).
\end{equation}
Note that, the individual values of $X_{11}$ and $X_{12}$ is often not required and the SSR can be performed from knowing the sum, $(X_{11}+X_{12})$ instead, retaining the blinding  and integrity of the trial.

In the presence of delay, though, the re-estimated sample size is impacted by the pipeline observations, as described previously in the main article in Section "Sample size re-estimation".
The final sample size $(N^*)$ for the trial can be obtained from Equation (2) of the main article.
The pipeline subjects ($n_\text{delay}$) are computed similar to Section "Computing the number of pipeline participants"
, where, the total recruitment time is assumed to be $t$ units (equal to 24 months in our simulations) to recruit all of $n_{init}$ patients determined in the planning stage.

\subsection{Results: Impact of delay on sample size re-estimation for a binary outcome variable}

For a binary outcome variable, we assumed that initially the sample size calculations are based on a control treatment success rate of $\pi_1 = p_1 = 0.3$ and to detect a treatment effect of $\delta_1 = 0.25$ (i.e., $\pi_2 = p_2 = 0.55$).
For this scenario, the trial requires a total of $n_0=94$ patients across both arms for a one-sided 5\% significance level and 80\% power.
The total recruitment time was assumed to be 24 months to recruit all 94 patients and patient accrual was based on uniform or linear recruitment.

Similar to the result section in the main article, three scenarios were investigated, where,
\begin{itemize}
    \item Case I: $\pi_1 = 0.1$.
    \item Case II: $\pi_1 = 0.3$. 
    \item Case III: $\pi_1 = 0.5$. 
\end{itemize} 
In each case, $\pi_2 = \pi_1 + \delta_1$.

We have considered varying $m = 0, 1, \dots, 24$ months.
For each parameter combination, 10,000 simulations were run to obtain the distribution of $N^*$. 
For each simulation, the first $n_1 = 30$ [approximately 33\% of $n_{init}$] samples across both arms were drawn from $Bin(1, \pi_1)$ and $Bin(1, \pi_2)$ populations, and the pooled sample success rate $(p)$ was computed, based on which the sample size was re-estimated.
The final sample size, $N^*$, was obtained through Equation (2) in the main article.
For the simulation purposes we have not imposed a maximum allowed sample size in order to observe the full distribution of $N^*$.

\subsubsection{Approach 1: Impact of delay on the distribution of re-estimated sample size}

First we plot the distribution of the final sample size post SSR for varying delay lengths, $m=0,3,6,\cdots, 24$.
Figure~\ref{fig5:N_SSR_uni_p0} and Figure~\ref{fig6:N_SSR_lin_p0} shows $N^*$ obtained through SSR for different $m$'s.
As observed for a continuous outcome, the spread of the distribution of $N^*$ reduces with increasing $m$.
This is due to the distribution being truncated at $N^* = n_1 + n_\text{delay}$.

Similar to the continuous case, the most severe impact of delay is observed for case I, when $\pi_1 = 0.1$.
Here, the required oracle sample size ($n_{oracle} = 66$) is less than the initially estimated required sample size ($n_{init} = 94$).
Therefore, the delay period often results in accruing more patients than are estimated as being required and the greater the delay is, the more pipeline patients will be contributing to the loss of efficiency.
Linearly increasing recruitment worsens the situation due to a higher number of pipeline patients.

\begin{figure}
    \centering
    \includegraphics[width=0.5\textwidth]{Figures/Final_sample_size_bin_uniform.jpeg}
    \caption{Distribution of the re-estimated sample size based on the decision to reject or accept the null for varying delay lengths, under uniformly recruited samples, for different values of $\pi_1$.}
    \label{fig5:N_SSR_uni_p0}
\end{figure}

\begin{figure}
    \centering
    \includegraphics[width=0.5\textwidth]{Figures/Final_sample_size_bin_linear.jpeg}
    \caption{Distribution of the final sample size based on the decision to reject or accept the null for varying delay lengths, under samples recruited at a linearly increasing rate, for different values of $\pi_1$.}
    \label{fig6:N_SSR_lin_p0}
\end{figure}

The \textit{delay impact} is also plotted in Figure~\ref{fig:ch4:delay_impact_SSR_binary} for the above simulation scenario.
It is interesting to note that the impact of delay on the considered trials with a binary outcome is more severe compared to the continuous outcome examples from earlier.
A reason for this may be the relatively smaller sample sizes required for the trials considered here compared to those in the normal outcome case.

\begin{figure}[htbp]
     \centering
     \begin{subfigure}[b]{0.5\textwidth}
         \centering
         \includegraphics[width=\textwidth]{Figures/delayimpact_bin_uniform.jpeg}
         \caption{Uniform recruitment.}
         \label{fig:ch4:Delay_impact_uni_bin}
     \end{subfigure}
     \hfill
     \begin{subfigure}[b]{0.5\textwidth}
         \centering
         \includegraphics[width=\textwidth]{Figures/delayimpact_bin_linear.jpeg}
         \caption{Linear recruitment.}
         \label{fig:ch4:Delay_impact_lin_bin}
     \end{subfigure}
        \caption{\textit{Delay impact} for varying delay lengths $(m = 1, 2, \dots, 24)$ for $\pi_1 = 0.1, 0.3, 0.5$, under uniform and linear recruitment patterns.}
        \label{fig:ch4:delay_impact_SSR_binary}
\end{figure}

Furthermore, it can be seen from Figure~\ref{fig6:N_SSR_lin_p0} that here $N^*$ takes a single value of 104 (here, $n_1=30$ and $n_{delay}=74$) for $m=18$.
Here, due to the recruited pipeline patients the final sample size already surpasses the the maximum sample size possible (100) for any $\pi \in (0,1)$ for $\alpha=0.05$ and $\beta=0.2$ (due to the binary nature of the data).
Thus for all simulation, the final sample size takes the constant value of 104, making \textit{delay impact= 1}.
Now if $m$ increases further, this value would increase rapidly with a higher value of $N^*$ consisting of $N^* = n_1 + n_\text{delay}$.
Thus we see for linear recruitment the \textit{delay impact} also quickly converges to 1 in Figure \ref{fig:ch4:delay_impact_SSR_binary}, where, for $m> 15$ the final sample size only takes the value $N^* = n_1 + n_\text{delay}$ with \textit{delay impact}= 1.

\subsubsection{Approach 3: Impact of delay on the cost metric}

Here, we plot the cost metric to understand the impact of delay on blinded SSR for a binary outcome variable.
Since similar inferences can be drawn from the cost metric plot alone, in this case, we have omitted the RMSE plot.

\begin{figure}
     \centering
     \begin{subfigure}[b]{1.1\textwidth}
         \centering
         \includegraphics[width=\textwidth]{Figures/cost_bin_uniform.jpeg}
         \caption{Uniform recruitment.}
         \label{fig7:cost_uni_p0}
     \end{subfigure}
     \hfill
     \begin{subfigure}[b]{1.1\textwidth}
         \centering
         \includegraphics[width=\textwidth]{Figures/cost_bin_linear.jpeg}
         \caption{Linear recruitment.}
         \label{fig7:cost_lin_p0}
     \end{subfigure}
        \caption{Cost for varying delay lengths for different values of $\pi_1 = 0.1$, 0.3 and 0.5 for uniform and linear recruitment patterns compared to a single stage design assuming $p_1 = 0.3$.}
        \label{fig:ch4:Cost_SSR_p0}
\end{figure}

It is evident from Figure~\ref{fig:ch4:Cost_SSR_p0} that an increase in delay length significantly impacts the cost of the trial.
It is also sensitive to the true value of the $\pi_i$. 
The trials where $\pi_1 = 0.1$ suffer higher costs in the presence of large delay ($m > 15$ months) due to a smaller oracle sample size, aligning with the previous inferences in the normal outcome case. 
Furthermore, also similar to prior observations, a linearly increasing recruitment rate increases the cost in comparison to uniformly recruited patients, as this results in a greater number of pipeline subjects.

The exact values for the \textit{delay impact}, cost, and MSE of the SSR designs with binary outcomes can be found in the following Tables~\ref{tab:SSR_uni_bin} and~\ref{tab:SSR_lin_bin}.

The following section contains the exact values of for \textit{delay impact}, $MSE$, the cost metric and empirical power values for different $m$ values (in particular, $m=0,3,6,\dots, 24$).
Table \ref{tab:SSR_uni_bin} provides these values along with the oracle sample size and average final sample sizes from the 10,000 simulation under a uniform recruitment assumption, whereas, table \ref{tab:SSR_lin_bin} provides the same for a linearly increasing recruitment pattern.

\begin{landscape}
\begin{table}[]
\caption{Impact of delay on Efficiency and cost parameters for a SSR design with uniform recruitment for a binary outcome variable.
Here, $n_{single}$ is computed based on $\pi_1=p_1=0.3, \delta_0=0.25$ and $\alpha=0.05, \beta=0.2$ and found to be 94 patients}
\label{tab:SSR_uni_bin}
\begin{tabular}{@{}cccccccccccc@{}}
\toprule
$\pi_1$ &
  \begin{tabular}[c]{@{}c@{}}Average\\ $p$\end{tabular} &
  \begin{tabular}[c]{@{}c@{}}Empirical \\ Power\end{tabular} &
  $m$ &
  $n_{oracle}$ &
  $n_{delay}$ &
  Avg. $N^*$ &
  $MSE_{single}$ &
  $MSE_{SSR}$ &
  $Cost_{single}$ &
  $Cost_{SSR}$ &
  Delay Impact \\ \midrule
0.1 & 0.216 & 0.793 & 0  & 66.872 & 0.000  & 67.699  & 767.704 & 238.101   & 8.457 & 3.381  & 0.019 \\
    & 0.217 & 0.796 & 3  &        & 11.822 & 68.246  &         & 208.562   &       & 2.734  & 0.061 \\
    & 0.219 & 0.815 & 6  &        & 23.645 & 69.439  &         & 156.880   &       & 1.892  & 0.155 \\
    & 0.219 & 0.834 & 9  &        & 35.467 & 72.943  &         & 112.741   &       & 1.280  & 0.471 \\
    & 0.221 & 0.874 & 12 &        & 47.290 & 79.403  &         & 175.656   &       & 2.020  & 0.652 \\
    & 0.222 & 0.904 & 15 &        & 59.112 & 89.341  &         & 506.203   &       & 5.668  & 0.952 \\
    & 0.224 & 0.933 & 18 &        & 70.935 & 100.935 &         & 1,160.258 &       & 12.566 & 1.000 \\
    & 0.223 & 0.955 & 21 &        & 82.757 & 112.757 &         & 2,105.435 &       & 22.256 & 1.000 \\
    & 0.224 & 0.968 & 24 &        & 94.580 & 124.580 &         & 3,330.153 &       & 34.607 & 1.000 \\ \midrule
0.3 & 0.424 & 0.797 & 0  & 94.580 & 0.000  & 94.554  & 0.000   & 45.680    & 0.000 & 0.631  & 0.000 \\
    & 0.424 & 0.799 & 3  &        & 11.822 & 94.460  &         & 47.407    &       & 0.654  & 0.000 \\
    & 0.423 & 0.795 & 6  &        & 23.645 & 94.548  &         & 46.588    &       & 0.637  & 0.000 \\
    & 0.423 & 0.793 & 9  &        & 35.467 & 94.462  &         & 44.753    &       & 0.602  & 0.007 \\
    & 0.423 & 0.793 & 12 &        & 47.290 & 94.708  &         & 37.761    &       & 0.494  & 0.022 \\
    & 0.423 & 0.800 & 15 &        & 59.112 & 95.702  &         & 18.386    &       & 0.229  & 0.198 \\
    & 0.420 & 0.821 & 18 &        & 70.935 & 100.935 &         & 40.387    &       & 0.491  & 1.000 \\
    & 0.422 & 0.861 & 21 &        & 82.757 & 112.757 &         & 330.423   &       & 3.847  & 1.000 \\
    & 0.423 & 0.885 & 24 &        & 94.580 & 124.580 &         & 900.000   &       & 10.133 & 1.000 \\ \midrule
0.5 & 0.628 & 0.802 & 0  & 90.622 & 0.000  & 90.553  & 15.665  & 87.060    & 0.192 & 1.223  & 0.000 \\
    & 0.628 & 0.798 & 3  &        & 11.822 & 90.597  &         & 83.520    &       & 1.165  & 0.000 \\
    & 0.627 & 0.798 & 6  &        & 23.645 & 90.697  &         & 81.926    &       & 1.124  & 0.003 \\
    & 0.627 & 0.795 & 9  &        & 35.467 & 90.775  &         & 76.166    &       & 1.015  & 0.028 \\
    & 0.626 & 0.808 & 12 &        & 47.290 & 91.324  &         & 55.382    &       & 0.705  & 0.066 \\
    & 0.622 & 0.811 & 15 &        & 59.112 & 93.646  &         & 27.930    &       & 0.337  & 0.392 \\
    & 0.619 & 0.847 & 18 &        & 70.935 & 100.935 &         & 106.357   &       & 1.271  & 1.000 \\
    & 0.621 & 0.867 & 21 &        & 82.757 & 112.757 &         & 489.975   &       & 5.620  & 1.000 \\
    & 0.623 & 0.901 & 24 &        & 94.580 & 124.580 &         & 1,153.135 &       & 12.815 & 1.000 \\ \bottomrule
\end{tabular}
\end{table}
\end{landscape}

\begin{landscape}
\begin{table}[]
\caption{Impact of delay on Efficiency and cost parameters for a SSR design with linear recruitment for a binary outcome variable.
Here, $n_{single}$ is computed based on $\pi_1=p_1=0.3, \delta_0=0.25$ and $\alpha=0.05, \beta=0.2$ and found to be 94 patients}
\label{tab:SSR_lin_bin}
\begin{tabular}{@{}clllllllllll@{}}
\toprule
$\pi_1$ &
  \begin{tabular}[c]{@{}l@{}}Average\\ $p$\end{tabular} &
  \begin{tabular}[c]{@{}l@{}}Empirical \\ Power\end{tabular} &
  $m$ &
  $n_{oracle}$ &
  $n_{delay}$ &
  Avg. $N^*$ &
  $MSE_{single}$ &
  $MSE_{SSR}$ &
  $Cost_{single}$ &
  $Cost_{SSR}$ &
  Delay Impact \\ \midrule
0.1 & 0.215 & 0.783 & 0  & 66.872 & 0.000   & 67.631  & 767.704 & 235.505   & 8.457 & 3.343   & 0.019 \\
    & 0.217 & 0.803 & 3  &        & 10.657  & 68.179  &         & 211.462   &       & 2.798   & 0.057 \\
    & 0.220 & 0.819 & 6  &        & 24.151  & 69.915  &         & 156.259   &       & 1.870   & 0.142 \\
    & 0.222 & 0.850 & 9  &        & 40.483  & 75.310  &         & 118.731   &       & 1.352   & 0.475 \\
    & 0.222 & 0.904 & 12 &        & 59.652  & 89.861  &         & 529.678   &       & 5.921   & 0.952 \\
    & 0.222 & 0.956 & 15 &        & 81.659  & 111.659 &         & 2005.827  &       & 21.243  & 1.000 \\
    & 0.224 & 0.975 & 18 &        & 106.503 & 136.503 &         & 4848.386  &       & 49.783  & 1.000 \\
    & 0.225 & 0.991 & 21 &        & 134.184 & 164.184 &         & 9469.555  &       & 95.733  & 1.000 \\
    & 0.224 & 0.996 & 24 &        & 164.703 & 194.703 &         & 16340.590 &       & 164.059 & 1.000 \\ \midrule
0.3 & 0.424 & 0.798 & 0  & 94.580 & 0.000   & 94.548  & 0.000   & 45.228    & 0.000 & 0.615   & 0.000 \\
    & 0.423 & 0.803 & 3  &        & 10.657  & 94.563  &         & 45.191    &       & 0.615   & 0.000 \\
    & 0.423 & 0.791 & 6  &        & 24.151  & 94.572  &         & 45.137    &       & 0.615   & 0.000 \\
    & 0.424 & 0.792 & 9  &        & 40.483  & 94.568  &         & 42.639    &       & 0.568   & 0.007 \\
    & 0.423 & 0.800 & 12 &        & 59.652  & 95.860  &         & 17.582    &       & 0.218   & 0.200 \\
    & 0.419 & 0.854 & 15 &        & 81.659  & 111.659 &         & 291.692   &       & 3.408   & 1.000 \\
    & 0.424 & 0.911 & 18 &        & 106.503 & 136.503 &         & 1757.528  &       & 19.270  & 1.000 \\
    & 0.424 & 0.947 & 21 &        & 134.184 & 164.184 &         & 4844.740  &       & 50.976  & 1.000 \\
    & 0.424 & 0.972 & 24 &        & 164.703 & 194.703 &         & 10024.583 &       & 102.899 & 1.000 \\ \midrule
0.5 & 0.628 & 0.797 & 0  & 90.622 & 0.000   & 90.679  & 15.665  & 85.139    & 0.192 & 1.192   & 0.000 \\
    & 0.629 & 0.796 & 3  &        & 10.657  & 90.558  &         & 86.006    &       & 1.205   & 0.001 \\
    & 0.628 & 0.791 & 6  &        & 24.151  & 90.665  &         & 83.666    &       & 1.148   & 0.003 \\
    & 0.628 & 0.800 & 9  &        & 40.483  & 90.884  &         & 70.184    &       & 0.919   & 0.027 \\
    & 0.621 & 0.808 & 12 &        & 59.652  & 93.804  &         & 26.793    &       & 0.323   & 0.390 \\
    & 0.615 & 0.858 & 15 &        & 81.659  & 111.659 &         & 442.548   &       & 5.093   & 1.000 \\
    & 0.623 & 0.920 & 18 &        & 106.503 & 136.503 &         & 2105.041  &       & 22.827  & 1.000 \\
    & 0.624 & 0.958 & 21 &        & 134.184 & 164.184 &         & 5411.369  &       & 56.518  & 1.000 \\
    & 0.623 & 0.979 & 24 &        & 164.703 & 194.703 &         & 10832.790 &       & 110.691 & 1.000 \\ \bottomrule
\end{tabular}
\end{table}
\end{landscape}

\bibliographystyle{abbrv}